# Quantum States of a Two-Level Atom trapped in a Helical Optical Tube

V. E. Lembessis, A. Lyras, and O. M. Aldossary

*Quantum Technology Group, Department of Physics and Astronomy,*

*College of Science, King Saud University, Riyadh 11451, Saudi Arabia*

**Abstract**

We investigate the quantized states of a two-level Rb atom which is trapped by the optical dipole potential when the atom interacts with a helical optical tube (HOT) light field. The analysis shows that the stationary states can be defined by a triplet of quantum numbers and the corresponding wavefunctions exhibit a twisted, spiral-like 3D spatial shape.

## I. Introduction

Optical vortices are light beams made up by photons which carry orbital angular momentum along their propagation direction. The most prominent members of this family are the so-called Laguerre-Gaussian (LG) beams and the Bessel beams. In such beams, the photon orbital angular momentum (OAM) is quantized and is given by $\ell\hbar$ where $\ell$ is a non-zero integer number known as the winding number of the beam [1]. These beams are also characterized by a rich transverse spatial intensity structure with a cylindrical symmetry around the propagation direction and zero on-axis intensity. The advent of optical vortex beams has given fresh impetus to different research fields such as mechanical effects of light on atoms and optical communications [2], [3].

Before the introduction of these laser beams, the mechanical effects of light were limited either in the cooling of the atomic motion by the so-called scattering force which acts along the propagation direction of the beam and emerges from the exchange of linear momentum between the photon and the atom, or in the trapping of the atom near the beam axis by the optical dipole force. The optical vortices brought new opportunities. The quantized orbital angular momentum of the photons is responsible for the generation of a torque along the beam axis superimposed on the atomic translational motion. This torque is responsible for the rotation of the atom around the beam axis [4]. The rich transverse intensity landscape with a cylindrically symmetric distribution of bright and dark regions offers new schemes for optical dipole trapping.

The most prominent optical dipole force scheme created by LG beams is the so-called optical Ferris wheel which is created by the interference of two co-propagating LG beams with opposite winding numbers (i.e. $\ell_1 = -\ell_2 = \ell$) [5]. The transverse intensity pattern of this coherent light field has a remarkable cylindrically symmetric pedal-like structure with $2|\ell|$ bright regions. This field is actually a two-dimensional optical lattice with cylindrical symmetry which enables new opportunities for atomic conveyors and realizations of cold atom collisional quantum gates [6]. If the two beams which make up the optical Ferris wheel are made to be counter-propagating then we can have a helical optical tube (HOT) [7]. This is a light field where the bright regions are in the form of twisted tubes in space. There are studies which suggest the utilization of this light field in quantum metrology and rotational sensing [8], [9]. The study of the classical motion of an atomic particle trapped in a rotating HOT showed that we can use it as an Archimedes' spiral for elevating atoms [10], [11]. Soon after, an experiment shown that this configuration can be used for the elevation of nanoparticles [12].

In this paper we study the quantized states of a very cold atom trapped in a HOT. We show the existence of wavefunctions that have a twisted wormhole-like spatial structure. The structure of the paper is as follows. In section II we quantize the translational atomic motion and we arrive at the corresponding Schrödinger equation. In section III we derive the full solution of the Schrödinger equation giving the eigenfunctions and the eigenergies of the atom trapped in the HOT. The paper is accompanied by an Appendix which presents analytically the calculations for the eigenfunctions.

## II. Quantization of the atomic motion

The light field of a HOT is created by the interference of two similar counter-propagating LG beams with opposite winding numbers $\ell_1 = -\ell_2 = \ell$. In our calculations we ignore the the Guoy phase and curvature effects since these effects become important for very tightly focused beams when the beam waist takes values equal to or smaller than the wavelength [2], .When the two beams are arranged to propagate along the z-axis the electric field of this configurations is given by:

$$\mathbf{E} = \frac{1}{2} E_0 u_{p,|\ell|}(\rho, z) \hat{\mathbf{e}}.$$

(1)

where:

$$u_{p,|\ell|}(\rho,z) = \frac{C_{p,|\ell|}}{\sqrt{1+z^2/z_R^2}} \left(\frac{\rho\sqrt{2}}{w(z)}\right)^{|\ell|} exp\left[-\frac{2\rho^2}{w_0^2}\right] L_p^{|\ell|}\left(\frac{2\rho^2}{w^2(z)}\right) cos(kz+\ell\phi),$$

(2)

with $w(z) = w_0\sqrt{1+(z^2/z_R^2)}$ the width of the two beams making up the HOT having a common beam waist $w_0$ and a Rayleigh range $z_R = \pi w_0^2/\lambda$, $\lambda$ the wavelength while $C_{p,|\ell|} = \sqrt{2p!/\pi(|\ell|+p)!}$. The vector $\hat{\mathbf{e}}$ defines the polarization of the Ferris wheel light field, linear or elliptical at the x-y plane. As it has been shown, [10-11], the dynamics of a two-level atom trapped in a far-detuned HOT light field, when $p = 0$, is determined by the following optical dipole potential:

$$U(r,\varphi,z) = \frac{\hbar\Omega_0^2}{\Delta} u_{|\ell|}^2(\rho,z) \cos^2(|\ell|\varphi + kz) \qquad (3)$$

where $\Omega_0$ is the Rabi frequency at the focus, associated with a Gaussian beam of the same power, beam waist and wavelength. In Fig. 1 we present a 3-D plot of the optical dipole potential.

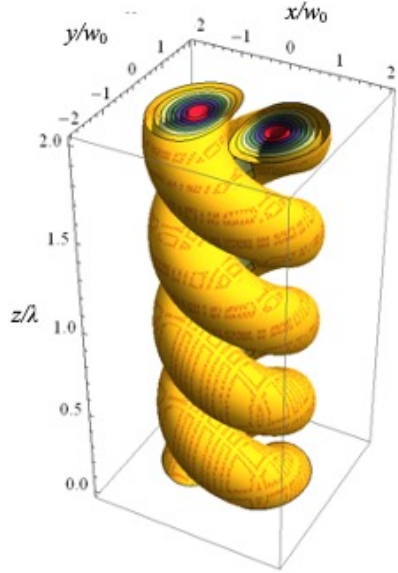

Figure1. Plot of the optical dipole potential in a HOT. Yellow color corresponds to the lowest values while red color to the highest ones.

Due to the helicoidal structure of the potential we choose to work in the so called Fresnet coordinates defined by:

$$\rho = r - \sqrt{\frac{|\ell|}{2}} w(z), \quad kv = \varphi + kz, \quad \xi = -\varphi = \frac{z}{h}, \qquad (4)$$

where $h = |\ell|/k$ is the helical pitch [7]. The classical motion of the atom is determined by its Langrangian which in the new coordinates assumes the form:

$$L = \frac{m}{2}\left\{\dot{r}^2 + (r^2 + h^2)\dot{\xi}^2 + \frac{r^2}{(r^2+h^2)}\dot{v}^2\right\} - U(\rho, \xi, v), \quad (5)$$

with

$$U(\rho, \xi, v) = \frac{\hbar \Omega_0^2}{\Delta} u_{|\ell|}^2(\rho, \xi) \cos^2(kv), \qquad (6)$$

and $u_{|\ell|}^2(\rho, \xi)$ is given as:

$$u_{|\ell|}^2(\rho, \xi) = C_{|\ell|}^2 \frac{w_0^2}{w^2(\xi)} \left(\frac{\sqrt{2}}{w(\xi)}\left(\sqrt{\frac{|\ell|}{2}} w(\xi) + \rho\right)\right)^{2l} exp\left(\frac{-2\left(\sqrt{\frac{|\ell|}{2}} w(\xi)+\rho\right)^2}{w^2(\xi)}\right), \quad (7)$$

with $w(\xi) = w_o\sqrt{\alpha \xi^2 + 1}$ and $\alpha = \frac{|l|^2 \lambda^4}{4\pi^4 w_0^4} = \left(\sqrt{\frac{|l|}{2}} \frac{\lambda}{\pi}\right)^4 \frac{1}{w_0^4}$. The classical study of the problem has shown that the radial motion is in the neighborhood of the maximum intensity points, thus $r \approx \sqrt{\frac{|\ell|}{2}} w(\xi)$ and $\dot{r} \approx \dot{\rho}$ with $4\rho^2 \ll w_o^2$ [10]. Moreover, it was shown that $z < z_R$ which gives $\alpha \xi^2 < 1$, and $kv \ll 1$. Under these conditions the Langrangian becomes:

$$L = \frac{m}{2}\left\{\dot{\rho}^2 + (r^2 + h^2)\dot{\xi}^2 + \frac{r^2}{(r^2+h^2)}\dot{v}^2\right\} - U(\rho, \xi, v) \qquad (8)$$

while the potential become:

$$U(\rho, \xi, v) = -\varepsilon + \frac{4\varepsilon \rho^2}{w_0^2} + \varepsilon \alpha \xi^2 + \varepsilon k^2 v^2 + \cdots \qquad (9)$$

where $\varepsilon$ is the depth of the dipole potential given by:

$$\varepsilon = \hbar \Omega_o^2 |l|^{|l|} e^{-|l|}/\Delta |l|! \quad (10)$$

which becomes shallower when the beam power decreases and when the detuning, beam waist, and winding number increase.

From the Langrangian we could get the Hamiltonian according to:

$$H = P_\rho \dot{\rho} + P_\xi \dot{\xi} + P_\nu \dot{\nu} - L. \quad (11)$$

The conjugate momenta which correspond to $\xi$, $\nu$, and $\rho$ are:

$$P_\xi = \frac{\partial L}{\partial \dot{\xi}} = m(r^2 + h^2)\dot{\xi} \quad \rightarrow \quad \dot{\xi} = \frac{P_\xi}{m(r^2+h^2)} \quad (12a)$$

$$P_\nu = \frac{\partial L}{\partial \dot{\nu}} = m\frac{r^2}{r^2+h^2}\dot{\nu} \quad \rightarrow \quad \dot{\nu} = \frac{r^2+h^2}{r^2}\frac{P_\nu}{m} \quad (12b)$$

$$P_\rho = \frac{\partial L}{\partial \dot{\rho}} = m\dot{\rho} \quad \rightarrow \quad \dot{\rho} = \frac{P_\rho}{m}. \quad (12c)$$

Let us define the parameters $M(\xi) = m(r^2 + h^2)$ and $D(\xi) = (r^2 + h^2)/r^2$. The quantized expressions of the above mentioned quantities are:

$$\dot{\xi} = \frac{1}{M(\xi)}\boldsymbol{P}_\xi, \quad \dot{\nu} = \frac{D(\xi)}{m}\boldsymbol{P}_\nu, \quad \dot{\rho} = \frac{1}{m}\boldsymbol{P}_\rho, \quad (13)$$

which leads to the quantized form of Langrangian given by:

$$\boldsymbol{L} = \frac{1}{2}\left\{\frac{1}{m}\boldsymbol{P}_\rho \cdot \boldsymbol{P}_\rho + \boldsymbol{P}_\xi \cdot \frac{1}{M(\xi)}\boldsymbol{P}_\xi + \frac{D(\xi)}{m}\boldsymbol{P}_\nu \cdot \boldsymbol{P}_\nu\right\} - U \quad (14)$$

Then substitute the Langrangian Eq. (14) into the Hamiltonian Eq. (11) to obtain the quantized Hamiltonian:

$$\boldsymbol{H} = \frac{1}{2m}\boldsymbol{P}_\rho \cdot \boldsymbol{P}_\rho + \frac{1}{2}\boldsymbol{P}_\xi \cdot \frac{1}{M(\xi)}\boldsymbol{P}_\xi + \frac{D(\xi)}{2m}\boldsymbol{P}_\nu \cdot \boldsymbol{P}_\nu + U, \quad (15)$$

where the momentum operators are:

$$\boldsymbol{P}_\rho = -i\hbar\frac{\partial}{\partial \rho}, \quad \boldsymbol{P}_\xi = -i\hbar\frac{\partial}{\partial \xi}, \quad \boldsymbol{P}_\nu = -i\hbar\frac{\partial}{\partial \nu} \quad (16)$$

The Hamiltonian should act on the wavefunction $\psi$ and lead to the following Schrödinger equation:

$$\left(\frac{-\hbar^2}{2m}\frac{\partial^2}{\partial \rho^2} + \frac{\partial}{\partial \xi}\frac{-\hbar^2}{2M(\xi)}\frac{\partial}{\partial \xi} + D(\xi)\frac{-\hbar^2}{2m}\frac{\partial^2}{\partial v^2} + U\right)\psi = E\psi \quad . \quad (17)$$

## II. Solution of the Schrödinger Equation

The Schrödinger equation can be simplified due to the fact that $h \ll w_o$, which is satisfied for small values of $|l|$ since $h = |\ell|/k = 2\pi\lambda|\ell|$. This condition ensures that $D(\xi) \approx 1$ and $M(\xi) \approx |l|mw^2(\xi)/2w_0^2 = |l|m(\alpha\xi^2 + 1)/2$. If we also recall that $\alpha\xi^2 < 1$ then we also have $1/2M(\xi) \approx 2(1 - \alpha\xi^2)/|l|m$. Then, the Schrödinger equation becomes as follows:

$$\left\{\frac{-\hbar^2}{2m}\frac{\partial^2}{\partial \rho^2} + \frac{\partial}{\partial \xi}\frac{-\hbar^2}{|l|mw_0^2}(1 - \alpha\xi^2)\frac{\partial}{\partial \xi} + \frac{-\hbar^2}{2m}\frac{\partial^2}{\partial v^2} + \frac{4\varepsilon\rho^2}{w_0^2} - \varepsilon(1 - \alpha\xi^2) + \varepsilon k^2 v^2\right\}\psi = E\psi \quad (18)$$

For the solution we are going to use the method of separation of variables. We assume that:

$$\psi = R(\rho)\Xi(\xi)N(v) \quad (19)$$

Then Eq. (18) becomes as follows:

$$\frac{-\hbar^2}{2mR(\rho)}\frac{d^2R(\rho)}{d\rho^2} + \frac{4\varepsilon\rho^2}{w_0^2} + \frac{1}{\Xi(\xi)}\frac{d}{d\xi}\frac{-\hbar^2}{|l|mw_0^2}(1 - \alpha\xi^2)\frac{d\Xi(\xi)}{d\xi}$$

$$-\varepsilon(1 - \alpha\xi^2) + \frac{-\hbar^2}{2mN(v)}\frac{d^2N(v)}{dv^2} + \varepsilon k^2 v^2 = E \quad (20)$$

From our classical treatment we have found that $\omega_\rho = \sqrt{8\varepsilon/mw_0^2}$ and $\omega_v = \sqrt{2\varepsilon k^2/m}$, then

$$\frac{-\hbar^2}{2mR(\rho)}\frac{d^2R(\rho)}{d\rho^2} + \frac{m\omega_\rho^2}{2}\rho^2 + \frac{1}{\Xi(\xi)}\frac{d}{d\xi}\frac{-\hbar^2}{lmw_0^2}(1 - \alpha\xi^2)\frac{d\Xi(\xi)}{d\xi} - \varepsilon(1 - \alpha\xi^2)$$

$$+ \frac{-\hbar^2}{2mN(v)}\frac{d^2N(v)}{dv^2} + \frac{m\omega_v^2}{2}v^2 = E \quad (21)$$

The first two terms in the left-hand side are function of $\rho$ only, so they should be equal to a constant $E_\rho$ :

$$-\frac{\hbar^2}{2m}\frac{d^2R}{d\rho^2} + \frac{m}{2}\omega_\rho^2\rho^2 R = E_\rho R \quad (22)$$

The last two terms in the left-hand side are function of $v$ only, so they should be equal to a constant $E_v$:

$$-\frac{\hbar^2}{2m}\frac{d^2 N}{dv^2} + \frac{m}{2}\omega_v^2 v^2 N = E_v N \quad (23)$$

Both Eqs. (22) and (23) are harmonic oscillator equations with well-known solutions. The eigen-energies are:

$$E_\rho = \hbar\omega_\rho\left(n_\rho + \frac{1}{2}\right), \quad n_\rho = 0,1,2,\ldots \quad (24)$$

$$E_v = \hbar\omega_v\left(n_v + \frac{1}{2}\right), \quad n_v = 0,1,2,\ldots \quad (25)$$

while the corresponding wavefunctions are:

$$R_{n_\rho}(\rho) = \frac{1}{\sqrt{2^{n_\rho} n_\rho!}}\left(\frac{m\omega_\rho}{\pi\hbar}\right)^{1/4} \exp\left\{-\frac{m\omega_\rho}{2\hbar}\rho^2\right\} H_{n_\rho}\left(\sqrt{\frac{m\omega_\rho}{\hbar}}\rho\right), \quad (26)$$

$$R_{n_v}(v) = \frac{1}{\sqrt{2^{n_v} n_v!}}\left(\frac{m\omega_v}{\pi\hbar}\right)^{1/4} \exp\left\{-\frac{m\omega_v}{2\hbar}v^2\right\} H_{n_v}\left(\sqrt{\frac{m\omega_v}{\hbar}}v\right), \quad (27)$$

where $H_n$ is the Hermite polynomial. This implies that Schrödinger equation Eq. (21) gets the following form:

$$\left(\frac{d}{d\xi}\frac{-\hbar^2}{|l|m w_0^2}(1-\alpha\xi^2)\frac{d}{d\xi} - \varepsilon(1-\alpha\xi^2)\right)\Xi(\xi) = E_\xi \Xi(\xi), \quad (28)$$

where $E = E_\rho + E_v + E_\xi$. The Schrödinger equation Eq. (28) is a position-dependent-mass Schrödinger equation. The solution of Eq. (28) is given in the Appendix I and has the following form:

$$\Xi_m(\xi) = \sqrt{\frac{2}{\pi}}\frac{1}{(1-\alpha\xi^2)^{\frac{1}{4}}}\begin{cases} ce_{2n}\left(\sin^{-1}(\sqrt{\alpha}\xi) + \frac{\pi}{2}, q\right) & m = 2n \\ se_{2n+2}\left(\sin^{-1}(\sqrt{\alpha}\xi) + \frac{\pi}{2}, q\right) & m = 2n+1 \end{cases}, \quad (29)$$

where the functions $ce_{2n}$ and $se_{2n+2}$ are the Mathieu functions of order $2n$ and $2n+2$ respectively [14]. The corresponding eigenvalues are:

$$E_m = -\varepsilon + \frac{\alpha^{\frac{3}{4}}}{2|\ell|}(4m+2)\sqrt{\varepsilon E_r} = -\varepsilon + \frac{\alpha^{\frac{3}{4}}}{2|\ell|}\sqrt{\varepsilon E_r}\begin{cases}(8n+2) & m=2n \\ (8n+6) & m=2n+1\end{cases}. \quad (n \geq 0) \quad (30)$$

The energy gap between consecutive energy levels is:

$$\Delta E = E_{m+1} - E_m = 2\frac{\alpha^{\frac{3}{4}}}{|\ell|}\sqrt{\varepsilon E_r}, \qquad (31)$$

while the number of allowed bound states in our potential is given by $N \approx |\ell|\sqrt{\varepsilon/E_r}/2\alpha^{\frac{3}{4}}$.

Our calculations have shown that both the energy step $\Delta E$ and the number of allowed stares $N$ becomes large when the beam power and detuning are changed to make the dipole potential deep. The dependence of $\Delta E$ and $N$ on the beam waist is more subtle. As the beam waist is increased, $\Delta E$ decreases while N increases. This behavior results from the combined scaling of $\alpha$ and $\varepsilon$ versus the beam waist. The parameter $\alpha$, that encodes the geometrical characteristics of the potential scales as $\alpha \sim w_0^{-4}$ while $\varepsilon$, the potential depth, as $\varepsilon \sim w_0^{-2}$. Then, it is straightforward to obtain from (31) that $\Delta E \sim w_0^{-4}$ while for N one obtains, from the corresponding expression, that $N \sim w_0^2$. Finally, the number of bound states N decreases when the winding number is increased, which amounts to a swallower potential depth. It is interesting, that $\Delta E$ also decreases with increasing winding number but at a slower rate compared to the potential depth. This results again from the combined scaling of $\alpha$ and $\varepsilon$ versus the winding number.

If we consider the transition $5^2S_{1/2} - 5^2P_{3/2}$ in a Rb atom and choose the following optical parameters: beam power $P_r=1mW$, detuning $\Delta = -2 \times 10^{15} Hz$, beam waist $w_0 = 30\mu m$, and $\ell = 1$, then $h/w_0 \approx 0.04$, and the approximations leading to Eq.(18) are well justified. In such a case, the $^{85}$Rb atom will be trapped in the dipole potential with depth $\varepsilon = 4.44E_r$, energy step $\Delta E \approx 2 \times 10^{-4}E_r$, and the number of bound states $N \approx 22030$. In Fig.2 we show the probability distribution $|\Xi_m(\xi)|^2$ of the first four states $\Xi_m(\xi)$.

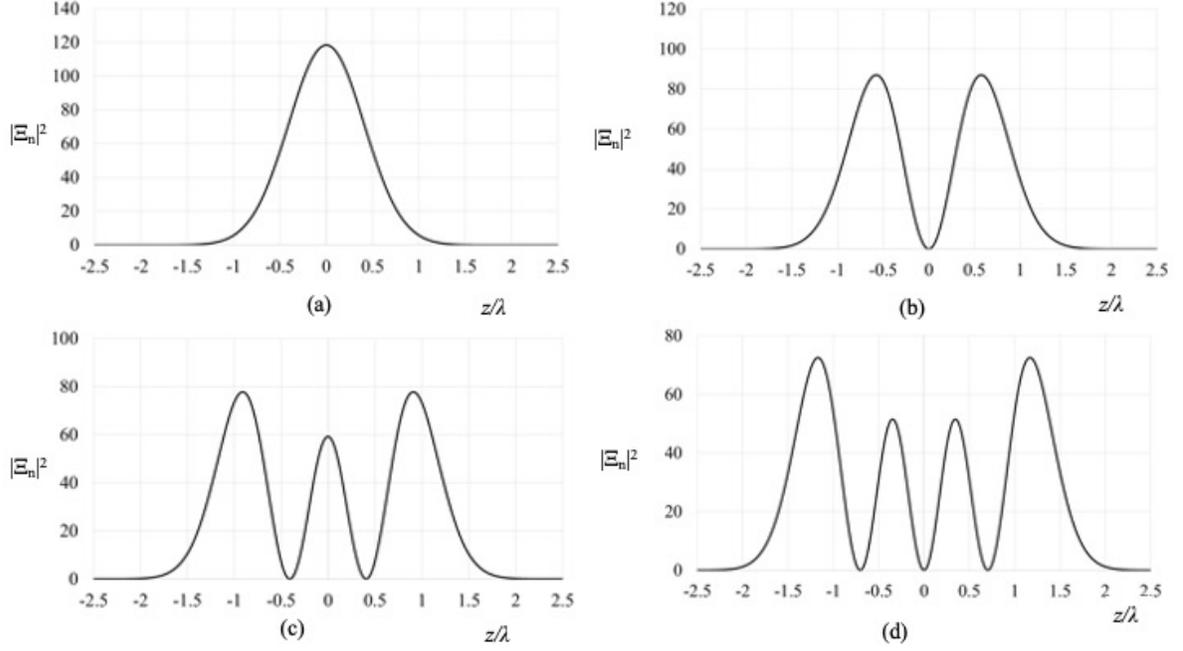

Figure 2. Plots of (a) $|\Xi_0(\xi)|^2$, (b) $|\Xi_1(\xi)|^2$, (c) $|\Xi_2(\xi)|^2$, and (d) $|\Xi_3(\xi)|^2$

The full expression of the wave function is:

$$\psi_{(n_\rho,n_\nu,n_\xi)}(r,\varphi,z) = R_{n_\rho}(r,z)N_{n_\nu}(\varphi,z)\Xi_{n_\xi}(z). \quad (32)$$

Where $R_{n_\rho}(x,y,z)$ is the wavefunction of radial mode $n_\rho$:

$$R_{n_\rho}(r,z) = \frac{1}{\sqrt{2^{n_\rho}n_\rho!}}\left(\frac{m\omega_\rho}{\pi\hbar}\right)^{1/4} \exp\left\{-\frac{m\omega_\rho}{2\hbar}[\rho(r,z)]^2\right\} H_{n_\rho}\left(\sqrt{\frac{m\omega_\rho}{\hbar}}\rho(r,z)\right), \quad (33)$$

with $\rho(r,z) = r - \sqrt{|\ell|/2}\,w(z)$. The quantity $N_{n_\nu}(\varphi,z)$ is the wavefunction of binormal mode $n_\nu$:

$$N_{n_\nu}(\varphi,z) = \frac{1}{\sqrt{2^{n_\nu}n_\nu!}}\left(\frac{m\omega_\nu}{\pi\hbar}\right)^{1/4} \exp\left\{-\frac{m\omega_\nu}{2\hbar}[\nu(\varphi,z)]^2\right\} H_{n_\nu}\left(\sqrt{\frac{m\omega_\nu}{\hbar}}\nu(\varphi,z)\right), \quad (34)$$

with $\nu(\varphi,z) = \ell\varphi + kz$. The quantity $\Xi_{n_\xi}(z)$ is the wavefunction of helical mode $n_\xi$:

$$\Xi_{n_\xi}(z) = \sqrt{\frac{2}{\pi}} \frac{1}{(1-\alpha\xi^2)^{\frac{1}{4}}} \begin{cases} ce_{2n}\left(\sin^{-1}\left(\frac{z}{z_R}\right)+\frac{\pi}{2},q\right) & n_\xi = 2n \\ se_{2n+2}\left(\sin^{-1}\left(\frac{z}{z_R}\right)+\frac{\pi}{2},q\right) & n_\xi = 2n+1 \end{cases} \quad (35)$$

The subscripts of the total wavefunction $(n_\rho, n_\nu, n_\xi)$ represent the state mode of the total wave function with total energy:

$$E_{(n_\rho,n_\nu,n_\xi)} = -\varepsilon + \hbar\omega_\rho\left(n_\rho+\frac{1}{2}\right) + \hbar\omega_\nu\left(n_\nu+\frac{1}{2}\right) + \frac{\alpha^{\frac{3}{4}}}{2l}(4n_\xi+2)\sqrt{\varepsilon E_r}. \quad (36)$$

The spatial probability distribution is given by $\left|\psi_{(n_\rho,n_\nu,n_\xi)}(r,\varphi,z)\right|^2 = \left|R_{n_\rho}(r,z)N_{n_\nu}(\varphi,z)\Xi_{n_\xi}(z)\right|^2$. Fig. 3 depicts the 3D plot and the x-y plots at different points along z of the probability distributions $\left|\psi_{(n_\rho,n_\nu,n_\xi)}(x,y,z)\right|^2$ of the ground state $(0,0,0)$ where the optical parameters are $P_r = 1mW$, $\Delta = -2 \times 10^{15} Hz$, $w_0 = 30\mu m$, $\ell = 1$, and $\varepsilon = 4.4 E_r$. In Figs. 3, 4, 5 and 6 we give 3D plots of some of the first excited trapped states.

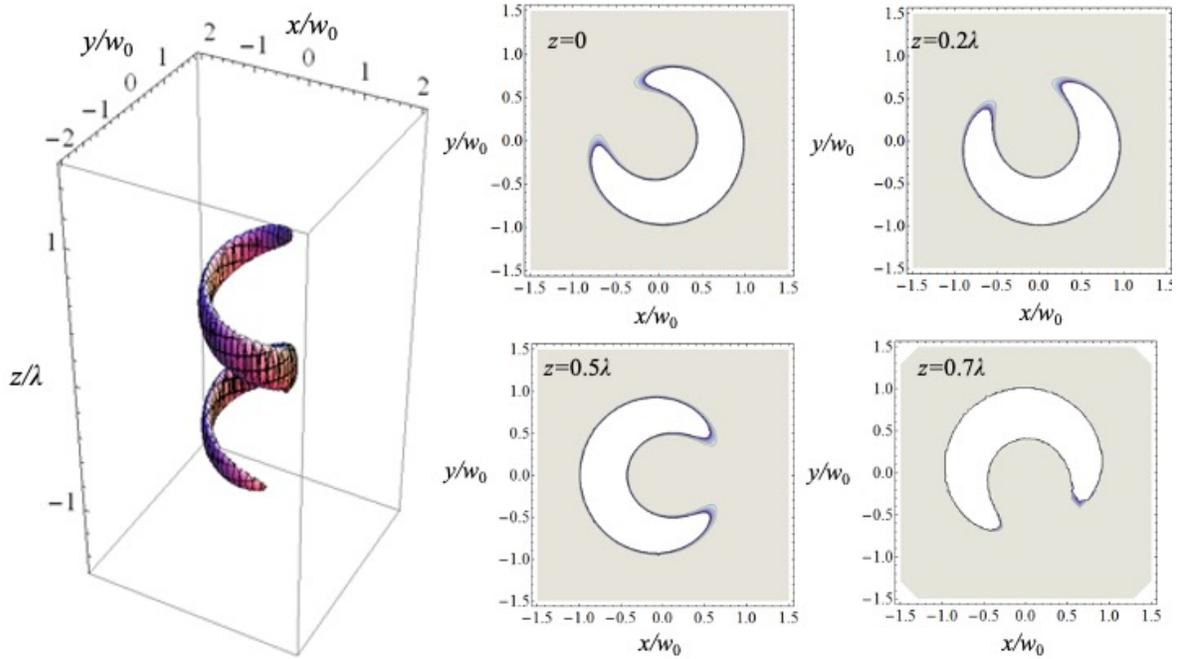

Figure 3. (a) 3D plot of the probability distribution at the ground state (0,0,0). (b) The probability distribution at the x-y plane for different axial positions. The optical parameters are: $P_r = 1mW$, $\Delta = -2 \times 10^{15} Hz$, $w_0 = 30 \mu m$, $\ell = 1$, and $\varepsilon = 4.4\, E_r$.

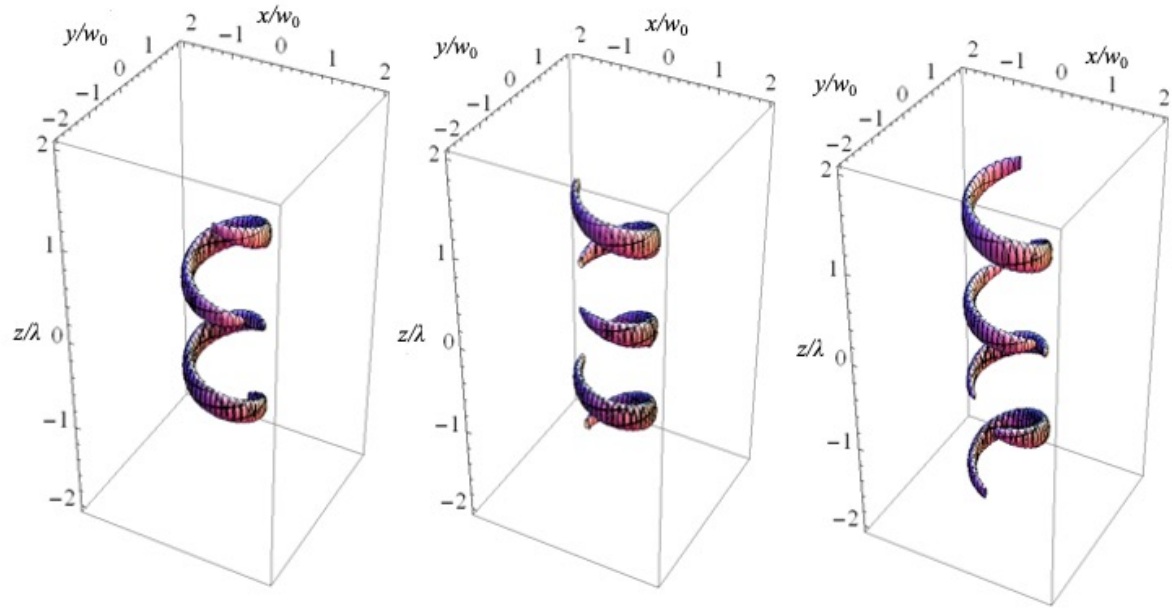

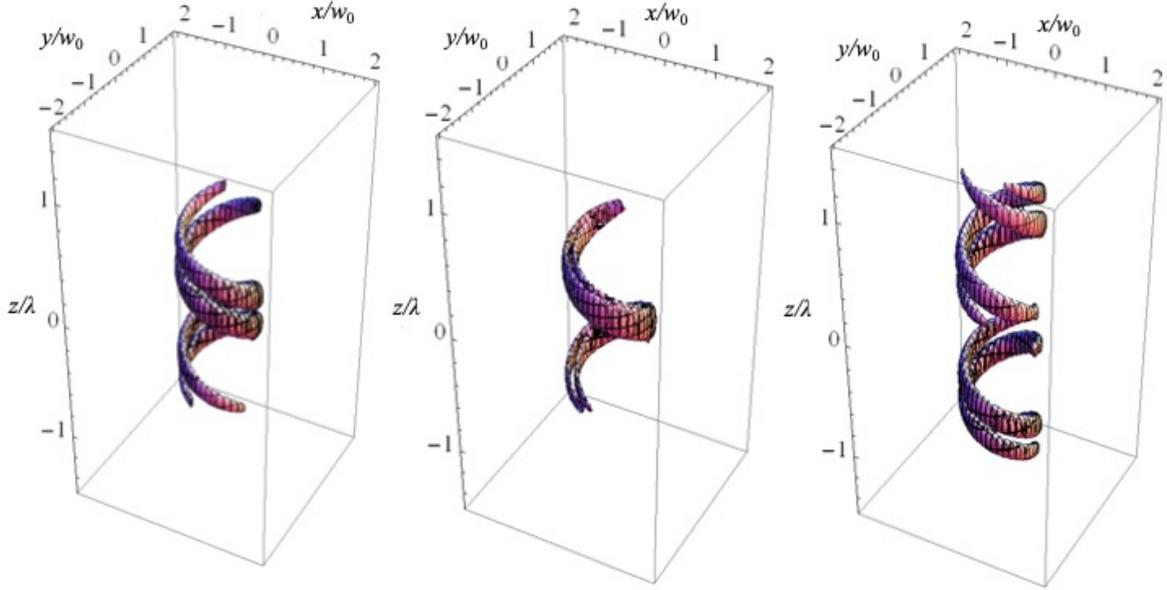

Figure 4. The 3D plots of the probability distribution for different modes (0,0,1), (0,0,2) and (0,0,3). The optical parameters are: $P_r = 1mW$, $\Delta = -2 \times 10^{15} Hz$, $w_0 = 30\mu m$, $\ell = 1$, and $\varepsilon = 4.4\, E_r$.

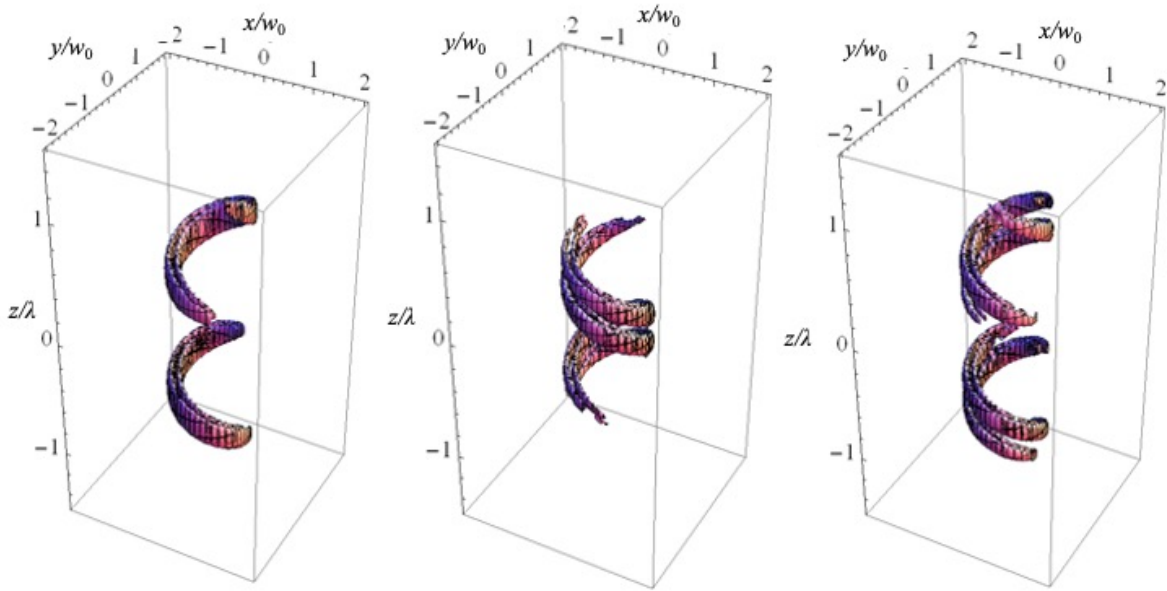

Figure 5. The 3D plots of the probability distribution for different modes (0,1,0), (1,0,0) and (0,1,1). The optical parameters are: $P_r = 1mW$, $\Delta = -2 \times 10^{15} Hz$, $w_0 = 30\mu m$, $\ell = 1$, and $\varepsilon = 4.4\, E_r$.

Figure 6. The 3D plots of the probability distribution for different modes (1,0,1), (1,1,0) and (1,1,1). The optical parameters are: $P_r = 1mW$, $\Delta = -2 \times 10^{15} Hz$, $w_0 = 30\mu m$, $\ell = 1$, and $\varepsilon = 4.4\ E_r$.

The calculations revealed, as evidenced in the figures 4 to 6, that the wavefunctions, and hence the spatial probability distribution, have a wormlike 3D shape that conforms with the spiral shape of the dipole potential (Fig. 1). An interesting nodal structure is revealed by the above figures (as well as others that have not been included for lack of space) supplemented by numerical data on the roots of the component wavefunctions $R_{n_\rho}(r,z)$, $N_{n_\nu}(\varphi,z)$, and $\Xi_{n_\xi}(z)$. We identify two classes of nodal structures; first, nodal lines that extend over the whole length of the probability distribution following a helical path, as in Fig. 5 (left and right panels) and Fig. 6 (middle and left panels); second, nodal gaps that are localized as shown in Fig. 4 (left and right panels), Fig. 5 (left panel) and Fig. 6 (right and left panels). The nodal lines are related to non-zero values of the $n_\nu$ quantum number, while the nodal gaps with non-zero values of the $n_\xi$ quantum number. It is revealing that the nodal lines are determined by the value of the $n_\nu$ quantum number, associated with the Frenet coordinate $\nu$, a combination of $\varphi$ and $z$. On the contrary, the nodal gaps are determined by the value of the $n_\xi$ quantum number, associated with $\varphi$ only. The number of nodal lines and nodal gaps increase as the corresponding quantum numbers increase, as evident in Fig. 4. A more detailed study of the nodal structure of the wavefunctions and its implications for the quantum dynamics of the trapped atoms will be presented in a future publication.

### III Conclusions and Outlook

We investigated the quantum energy eigevalues and the corresponding eigenfunctions of a cold two-level atom trapped by the optical dipole potential generated from the interaction of the atom with the far detuned light field of a HOT. To the best of our knowledge it is the first time in the literature that this problem is studied. We showed that the bound states are described by wavefunctions that have a spiral spatial structure. We performed numerical calculations employing values for the relevant parameters that are close to those used in atom trapping experiments assuming that our two-level system is a Rb atom. We gave plots of the spatial probability density distribution for the atomic ground state and some of the first excited states. We briefly discussed the spatial structure of these plots and qualitatively interpreted the relation between the values of the three quantum numbers defining each stationary state and the corresponding spatial features of the wavefunction and the probability density.

We can envisage a number of applications for the optical dipole potential studied in this work. At first, it could be employed as an alternative platform for single-atom dipole trapping that has previously been demonstrated for both Rb and Cs atoms using the typical optical dipole trap setup [17, 18]. Furthermore, it could be combined with Raman sideband cooling to trap a single atom in the quantum ground state of the potential, a process already demonstrated for a Rb atom in an optical tweezer [19]. Such studies, both theoretical and experimental, would one the one hand provide further insight into the properties of the quantum states of the HOT optical dipole potential while on the other hand exploit these properties for quantum information processing and quantum simulation schemes.

**Acknowledgment.** This project was funded by the National Plan for Science, Technology and Innovation (MAARIFAH), King Abdulaziz City for Science and Technology, Kingdom of Saudi Arabia, Award Number (15-MAT5110-02). We also thank Dr. A. Alrsheed for his valuable comments and suggestions on the calculations of the paper.

**Disclosures.**

The authors declare no conflicts of interest.

**Data Availability.** Data underlying the results presented in this paper are not publicly available at this time but may be obtained from the authors upon reasonable request.

**Appendix I:**

We work here on the solution of Eq. (28). We assume the following dimensionless parameters:

$$V(\xi) = -\frac{lmw_0^2\varepsilon}{\hbar^2}(1-\alpha\xi^2) = -\tilde{\varepsilon}(1-\alpha\xi^2) \quad \text{(AI.1a)}$$

$$\mathcal{M}(\xi) = \frac{1}{1-\alpha\xi^2}, \quad \text{(AI.1b)}$$

$$\tilde{E}_n = \frac{lmw_0^2 E_\xi}{\hbar^2} \quad \text{(AI.1c)}$$

where $\tilde{\varepsilon} = \frac{lmw_0^2\varepsilon}{\hbar^2}$.

Then Eq. (28) becomes as follows:

$$\left(\frac{1}{\mathcal{M}(\xi)}\frac{d^2}{d\xi^2} + \left(\frac{1}{\mathcal{M}(\xi)}\right)'\frac{d}{d\xi} + [-V(\xi) + \tilde{E}_n]\right)\Xi(\xi) = 0. \quad \text{(AI.2)}$$

Using the so-called point canonical transformation, [14], $\xi = F(u) = g^{-1}(u)$ where:

$$u = g(\xi) = \int_0^\xi \sqrt{\mathcal{M}(t)}\,dt \quad \text{(AI.3)}$$

The wavefunction is also transformed as:

$$\Xi(F(u)) = \varphi_n(u)e^{-\int_0^u W(t)dt}, \quad \text{(AI.4)}$$

where $W(u) = d\ln\left(\mathcal{M}(F(u))\right)^{-1/4}/du$.

Then the position-dependent mass Schrödinger equation Eq. (AI.2) transforms to a Schrödinger-like equation:

$$-\varphi_n''(u) + V_{eff}(u)\varphi_n(u) = \tilde{E}_n\varphi_n(u) \quad \text{(AI.5)}$$

where $V_{eff}(u)$ is the effective potential:

$$V_{eff}(u) = V(F(u)) + W^2(u) + W'(u). \quad \text{(AI.6)}$$

Since $\mathcal{M}(t) = 1/(1-\alpha\xi^2)$, then

$$u = \frac{1}{\sqrt{\alpha}}\int_0^{\sqrt{\alpha}\xi}\sqrt{\frac{1}{1-x^2}}\,dx = \frac{1}{\sqrt{\alpha}}\sin^{-1}(\sqrt{\alpha}\xi). \quad \text{(AI.7)}$$

The quantity $\sqrt{\alpha}\xi$ can be written as

$$\sqrt{\alpha}\xi = \sin(\sqrt{\alpha}u), \qquad (AI.8)$$

so the quantity $\mathcal{M}(F(u))$ can be expressed as

$$\mathcal{M}(F(u)) = \sec^2(\sqrt{\alpha}u). \qquad (AI.9)$$

Then, $W(u)$ becomes as follows:

$$W(u) = -\frac{\sqrt{\alpha}}{2}\tan(\sqrt{\alpha}u) \qquad (AI.10)$$

The effective potential given by Eq. (AI.6) becomes as follows:

$$V_{eff}(u) = -\tilde{\varepsilon}\cos^2(\sqrt{\alpha}u) - \frac{\alpha}{4}\left(\sec^2(\sqrt{\alpha}u) + 1\right). \qquad (AI.11)$$

But $\tilde{\varepsilon} \gg \alpha/4$ (for our parameters $\tilde{\varepsilon} \approx 3372$ and $\alpha \approx 10^{-6}$), then the effective is simplified as:

$$V_{eff}(u) = -\tilde{\varepsilon}\cos^2(\sqrt{\alpha}u). \qquad (AI.12)$$

Finally, the Schrödinger-like equation that is given by Eq. (AI.5) gets the form:

$$\varphi''(u) + \tilde{\varepsilon}\cos^2(\sqrt{\alpha}u)\,\varphi(u) = -\tilde{E}_n\varphi(u). \qquad (AI.13)$$

Since $-1 < \sin(\sqrt{\alpha}u) < 1$, then the quantity $\sqrt{\alpha}u$ gets values in the region $-\frac{\pi}{2} < \sqrt{\alpha}u < \frac{\pi}{2}$. Let $\sqrt{\alpha}u = x - \pi/2$, then $\cos(x - \pi/2) = \sin(x)$, where $0 < x < \pi$. The second derivatives $\varphi''(u)$ in Eq. (AI.13) becomes as $\alpha\varphi''(x)$. Then Eq. (AI.13) becomes:

$$\varphi''(x) + \frac{\tilde{\varepsilon}}{\alpha}\sin^2(x)\,\varphi(x) = -\frac{\tilde{E}_n}{\alpha}\varphi(x) \qquad (AI.14)$$

Using $\sin^2(x) = [1 - \cos(2x)]/2$, Eq. (AI.14) can be written as:

$$\varphi''(x) + \left[\frac{\tilde{E}_n}{\alpha} + \frac{\tilde{\varepsilon}}{2\alpha} - \frac{\tilde{\varepsilon}}{2\alpha}\cos(2x)\right]\varphi(x) = 0. \qquad (AI.15)$$

Let's use the following definitions:

$$a = \frac{\tilde{E}_n}{\alpha} + \frac{\tilde{\varepsilon}}{2\alpha}, \quad q = \frac{\tilde{\varepsilon}}{4\alpha} \qquad (AI.16)$$

Then Eq. (AI.15) becomes a Mathieu differential equation [14]:

$$\varphi''(x) + [a - 2q\cos(2x)]\varphi(x) = 0, \quad (AI.17)$$

where $a$ represents the characteristic values (eigenvalues) of the equation. The general solution of Mathieu differential equation is [15]:

$$\varphi(x) = \sum_{n=1}^{\infty}\{c_n ce_n(x,q) + d_n se_n(x,q)\}. \quad (AI.18)$$

Mathieu functions can be expanded in Fourier series with real coefficients with their corresponding characteristic values (or eigenvalues) for $n \geq 0$ as follows [16]:

$$ce_{2n}(x,q) = \sum_{r=0}^{\infty} A_{2r}^{(2n)}(q)\cos(2rx) \quad , \quad [a_{2n}(q)] \quad (AI.19a)$$

$$se_{2n+1}(x,q) = \sum_{r=0}^{\infty} B_{2r+1}^{(2n+1)}(q)\sin((2r+1)x), \quad [b_{2n+1}(q)] \quad (AI.19b)$$

$$ce_{2n+1}(x,q) = \sum_{r=0}^{\infty} A_{2r+1}^{(2n+1)}(q)\cos((2r+1)x), \quad [a_{2n+1}(q)] \quad (AI.19c)$$

$$se_{2n+2}(x,q) = \sum_{r=0}^{\infty} B_{2r+2}^{(2n+2)}(q)\sin((2r+2)x), \quad [b_{2n+2}(q)] \quad (AI.19d)$$

The periodicity of the first and last eigenfunctions is $\pi$ while the periodicity of the second and third eigenfunctions is $2\pi$ [16]. However, the periodicity of our problem is $\pi$ since $0 < x < \pi$. Therefore, the acceptable solutions, for integer values of $n \geq 0$, in our case are:

$$ce_{2n}(x,q) = \sum_{r=0}^{\infty} A_{2r}^{(2n)}(q)\cos(2rx), \quad [a_{2n}(q)] \quad (AI.20)$$

$$se_{2(n+1)}(x,q) = \sum_{r=0}^{\infty} B_{2(r+1)}^{(2(n+1))}(q)\sin(2(r+1)x), \quad [b_{2n+2}(q)]. \quad (AI.21)$$

Mathieu functions are orthogonal [16], thus,

$$\int_0^{\pi} se_{2n}(x,q)\, ce_{2m}(x,q)dx = 0 \quad (AI.22a)$$

$$\int_0^{\pi} se_{2n}(x,q)\, se_{2m}(x,q)dx = \int_0^{\pi} ce_{2n}(x,q)\, ce_{2m}(x,q)dx = \frac{\pi}{2}\delta_{nm} \quad (AI.22b)$$

From Eq. (AI16) we can deduce the eigen-energies of the system:

$$\begin{cases} \tilde{E}_{2n} \\ \tilde{E}_{2n+1} \end{cases} = -\frac{\tilde{\varepsilon}}{2} + \alpha \begin{cases} a_{2n}(q) \\ b_{2n+2}(q) \end{cases} = -2\alpha q + \alpha \begin{cases} a_{2n}(q) \\ b_{2n+2}(q) \end{cases}. \qquad \text{(AI.23)}$$

In our case $q > 0$, then the characteristic values are ordered is $a_{2n}(q) < b_{2n+2}$ $(a_0(q) < b_2(q) < a_2(q) < b_4(q) < a_4(q) < b_6(q) < a_6(q))$.

Using the definition $\tilde{\varepsilon}(= |\ell|mw_0^2\varepsilon/\hbar^2)$, the recoil energy

$$E_r = \frac{\hbar^2 k^2}{2m} \qquad \text{(AI.24)}$$

and the definition of $q$ as in Eq. (55b), we can write:

$$q = \frac{\tilde{\varepsilon}}{4\alpha} = \frac{|\ell|^2}{4\alpha^{3/2}} \frac{\varepsilon}{E_r} \qquad \text{(AI.25)}$$

Our numerical analysis for $q$ as a function of beam waist and for different values of power and detuning has shown that the values $q$ are sufficiently large. The characteristic can be expanded in terms of $q$ (for large values) as [13]:

$$a_r(q) \approx b_{r+1}(q) = -2q + (4r+2)\sqrt{q} - \frac{1}{2}\left(r^2 + r + \frac{1}{2}\right) + O\left(q^{-\frac{1}{2}}\right) \quad \text{(AI.26)}$$

Since $q$ is sufficiently large then the asymptotic characteristic values become:

$$a_{2n}(q) \approx -2q + (8n+2)\sqrt{q}, \quad b_{2n+2}(q) \approx -2q + (8n+6)\sqrt{q}, \qquad \text{(AI.27)}$$

which can be summarized as

$$a_m(q) \approx \begin{cases} a_m(q) & m = 2n \\ b_{m+1}(q) & m = 2n+1 \end{cases} = -2q + (4m+2)\sqrt{q}. \qquad \text{(AI.28)}$$

From Eq. (AI.23) we obtain the eigen-energy:

$$\tilde{E}_m = -4\alpha q + \alpha(4m+2)\sqrt{q}, \qquad \text{(AI.29)}$$

where $= 0,1,2,\cdots$. The corresponding eigenfunction for $m = 2n$ is:

$$\varphi_{2n} = c_{2n} ce_{2n}(x,q) \qquad \text{(AI.30)}$$

For $m = 2n + 1$ it is:

$$\varphi_{2n+1} = d_{2n+1} se_{2n+2}(x, q) \qquad \text{(AI.31)}$$

Using Eq. (AI.1c), Eq. (AI.24), and Eq. (AI.25) to write the eigenenergy as:

$$E_m = -\varepsilon + \frac{\alpha^{\frac{3}{4}}}{2l}(4m + 2)\sqrt{\varepsilon E_r} \qquad \text{(AI.32)}$$

the corresponding eigenstates have the following form:

$$\varphi_m(x) = \begin{cases} c_m ce_m(x, q) & m = 2n \\ d_m se_{m+1}(x, q) & m = 2n + 1 \end{cases} \qquad \text{(AI.33)}$$

The eigenfunction of our original equation Eq. (AI.4) is:

$$\Xi(F(u)) = \varphi_m(u) e^{-\int_0^u W(t) dt} \qquad \text{(AI.34)}$$

This can be written as (see Appendix II) as:

$$\Xi(\xi) = \varphi_m(g(\xi))\sqrt{g'(\xi)}. \qquad \text{(AI.35)}$$

Since $u = g(\xi)$ then it is easy to prove that if $\varphi_m(u)$ is normalized then $\Xi(x)$ is also normalized (see Appendix II), $\int |\Xi(\xi)|^2 d\xi = \int |\varphi_m(u)|^2 du = 1$. We use Eq. (AI.10) to find the following result:

$$\int_0^u W(t) dt = -\frac{\sqrt{\alpha}}{2} \int_0^u \tan(\sqrt{\alpha} t) dt = \ln\left(\sqrt{|\cos\sqrt{\alpha}u|}\right). \qquad \text{(AI.36)}$$

Then the eigenfunction in Eq. (AI.34) becomes:

$$\Xi_m(F(u)) = \frac{\varphi_m(u)}{\sqrt{|\cos\sqrt{\alpha}u|}}. \qquad \text{(AI.37)}$$

Since $\sqrt{\alpha}u = \sin^{-1}(\sqrt{\alpha}\xi)$ then $\cos\sqrt{\alpha}u = \sqrt{1 - \alpha\xi^2}$. Moreover, $\varphi_m$ in Eq. (AI.33) is a function of $x = \sqrt{\alpha}u + \pi/2 = \sin^{-1}(\sqrt{\alpha}\xi) + \pi/2$, which can be written as:

$$\varphi_m(\xi) = \begin{cases} c_m ce_m\left(\sin^{-1}(\sqrt{\alpha}\xi) + \frac{\pi}{2}, q\right) & m = 2n \\ d_m se_{m+1}\left(\sin^{-1}(\sqrt{\alpha}\xi) + \frac{\pi}{2}, q\right) & m = 2n + 1 \end{cases} \qquad \text{(AI.38)}$$

Using the normalization conditions $\int |\Xi(\xi)|^2 d\xi = 1$ we finally get:

$$\Xi_m(\xi) = \sqrt{\frac{2}{\pi}} \frac{1}{(1-\alpha\xi^2)^{\frac{1}{4}}} \begin{cases} ce_{2n}\left(\sin^{-1}(\sqrt{\alpha}\xi) + \frac{\pi}{2}, q\right) & m = 2n \\ se_{2n+2}\left(\sin^{-1}(\sqrt{\alpha}\xi) + \frac{\pi}{2}, q\right) & m = 2n+1 \end{cases} \quad (AI.39)$$

The corresponding eigenvalues are:

$$E_m = -\varepsilon + \frac{\alpha^{\frac{3}{4}}}{2l}(4m+2)\sqrt{\varepsilon E_r} = -\varepsilon + \frac{\alpha^{\frac{3}{4}}}{2l}\sqrt{\varepsilon E_r}\begin{cases}(8n+2) & m = 2n \\ (8n+6) & m = 2n+1\end{cases} \quad (AI.40)$$

**Appendix II: Proof of Eq. (AI.35).**

As we have seen the quantity $W(u)$ is defined (right after Eq. (AI.4)) as

$$W(u) = \frac{d\ln\big(M(F(u))\big)^{-1/4}}{du}. \qquad (AII.1)$$

Substituting in the following integration we get:

$$\int_0^u W(t)dt = \int_0^u \frac{d\ln\big(M(F(t))\big)^{-1/4}}{dt}dt = \ln\big(M(F(u))\big)^{-1/4}. \qquad (AII.2)$$

Then Eq. (AI.34) can be simplified as follows:

$$\Xi(\xi = F(u)) = \varphi_n(u)e^{-\int_0^u W(t)dt} = \varphi_n(u)e^{-\ln\big(M(F(u))\big)^{-1/4}}$$

$$= \varphi_n(u)e^{\ln\big(M(F(u))\big)^{1/4}} = \varphi_n(u)\big(M(F(u))\big)^{1/4}. \qquad (AII.3)$$

But $u = g(\xi) = \int_0^\xi \sqrt{M(t)}\,dt$, then $g'(\xi) = \sqrt{M(\xi)}$ to get finally:

$$\Xi(\xi) = \varphi_n(g(\xi))\sqrt{g'(\xi)}\ . \qquad (AII.4)$$

The normalization condition implies:

$$\int |\Xi(\xi)|^2 d\xi = \int |\varphi_n(g(\xi))|^2 g'(\xi)d\xi = \int |\varphi_n(g(\xi))|^2 \frac{dg(\xi)}{d\xi}d\xi = \int |\varphi_n(g(\xi))|^2 dg(\xi)$$

$$(AII.5)$$

Using $u = g(\xi)$, then (AII.5) becomes:

$$\int |\Xi(\xi)|^2 d\xi = \int |\varphi_n(u)|^2 du = 1 \qquad (AII.6)$$